\documentclass[superscriptaddress,prb,12pt,longbibliography]{revtex4-1}
\usepackage{graphicx}
\usepackage{hyperref}
\usepackage{float}
\usepackage{color}

\begin{document}
	
	\title{Soft X-ray spectro-ptychography on boron nitride nanotubes, carbon nanotubes and permalloy nanorods}
	
	\author{Jaianth Vijayakumar}
	\thanks{cptjaianth@gmail.com}
	\thanks{Present address: University of Ghent Center for X-ray tomography, N21 Proeftuinstraat, 9000 Ghent, Belgium}
	\affiliation{Synchrotron SOLEIL, L’Orme des Merisiers, Saint-Aubin, BP 48, 91192 Gif-sur-Yvette Cedex, France}
	\author{Hao Yuan}
	\affiliation{Department of Chemistry \& Chemical Biology, McMaster University, Hamilton, ON, Canada, L8S 4M1}
	\author{Nicolas Mille}
	\affiliation{Synchrotron SOLEIL, L’Orme des Merisiers, Saint-Aubin, BP 48, 91192 Gif-sur-Yvette Cedex, France}
	\author{Stefan Stanescu}
	\affiliation{Synchrotron SOLEIL, L’Orme des Merisiers, Saint-Aubin, BP 48, 91192 Gif-sur-Yvette Cedex, France}
	\author{Sufal Swaraj}
	\affiliation{Synchrotron SOLEIL, L’Orme des Merisiers, Saint-Aubin, BP 48, 91192 Gif-sur-Yvette Cedex, France}
	
	\author{Vincent Favre-Nicolin}
	\affiliation{ESRF, The European Synchrotron, 71 Avenue des Martyrs, 38000 Grenoble, France}
	\affiliation{Université Grenoble Alpes, Grenoble, France}
	\author{Ebrahim Najafi}
	\affiliation{The Chemours Company, Wilmington, Delaware 19899 US}
	\author{Adam Hitchcock}
	\affiliation{Department of Chemistry \& Chemical Biology, McMaster University, Hamilton, ON, Canada, L8S 4M1}
	\author{Rachid Belkhou}
	\affiliation{Synchrotron SOLEIL, L’Orme des Merisiers, Saint-Aubin, BP 48, 91192 Gif-sur-Yvette Cedex, France}


\begin{abstract}
Spectro-ptychography offers improved spatial resolution and additional phase spectral information relative to that provided by scanning transmission X-ray microscopes (STXM). However, carrying out ptychography at the lower range of soft X-ray energies (e.g., below 200 eV to 600 eV) on samples with weakly scattering signals can be challenging. We present soft X-ray ptychography results at energies as low as 180 eV and illustrate the capabilities with results from permalloy nanorods (Fe 2p), carbon nanotubes (C 1s), and boron nitride bamboo nanostructures (B 1s, N1s). We describe optimization of low energy X-ray spectro-ptychography and discuss important challenges associated with measurement approaches, reconstruction algorithms, and their effects on the reconstructed images. A method for evaluating the increase in radiation dose when using overlapping sampling is presented.
\end{abstract}
\maketitle

\section{Introduction}

The increasing availability of high flux of coherent X-rays at 3$^{\textrm{rd}}$ and 4$^{\textrm{th}}$ generation synchrotron light sources and X-ray free electron lasers has led to greater use of coherence enhanced microscopies, such as coherent diffraction imaging (CDI) and ptychography.~\cite{F_Pfeiffer_2018} Ptychography is an extension of scanning transmission X-ray microscopy (STXM) which can provide better spatial resolution than conventional STXM.~\cite{Adam_2015} Ultimately, it is expected that a fully optimized source, beamline optics and STXM/ptychography instrument will lead to an ultimate spatial resolution given by the diffraction limit at the X-ray wavelength used ($\lambda$/2). In STXM, the integrated flux of X-rays transmitted through a raster scanned sample is used to generate images. Sequences of such images, measured over a range of photon energies and photon polarizations can be analyzed to reveal the spatially resolved chemical, electronic and geometric or magnetic orientational properties of the sample.~\cite{Adam_2015} The spatial resolution of a conventional STXM is typically around 30 nm in current instruments, and depends on the properties of the Fresnel zone plate focusing optics, precision of scanning stages, mechanical stability and X-ray beam stability. The ptychography extension of STXM involves replacing the single channel integrating detector with a 2D detector (camera), which is used to measure the coherent scattering/diffraction~\cite{Edo_2013} of the transmitted X-rays in the far field. A set of such images measured at positions where the spots overlap is then reconstructed using computational algorithms to form amplitude and phase images of the sample and the X-ray probe at the sample. \cite{Finenup_1982,M_Dierolf_2010, F_Pfeiffer_2018, Rodenburg_2019, A_M_Maiden_2013} 

Ptychographic reconstruction enhances spatial resolution relative to that achieved by a STXM measurement using the same zone plate, with full focus. Until now soft X-ray ptychography has been carried out on synthetic~\cite{X_Shi_2016, D_A_Shapiro_2017, CP_Wang_2017, W_Li_2019, J_Grafe_2020} and biologically generated~\cite{X_Zhu_2016} magnetic materials, fuel cells,~\cite{J_Wu_2018} catalyst particles,~\cite{A_M_Wise_2016} lithium battery materials,~\cite{Y_S_Yu_2015, Y_S_Yu_2018, D_Shapiro_2014, T_Sun_2020} nanocomposites,~\cite{H_Yuan_2021, B_Bozzini_2017} cement materials (calcium silicate hydrates),~\cite{S_Bae_2015} diatoms,~\cite{M_Rose_2015, K_Giewe_2011} fibroblast cells,~\cite{M_Rose_2018} organic thin films,~\cite{V_Savikhin_2019} and carbon nanotubes.~\cite{N_Mille_2022} Prior to the development of the HERMES low energy ptychography capability,~\cite{N_Mille_2022} synchrotron-based ptychography at X-ray energies below 500 eV had not been reported, mainly due to limitations of the cameras used. In particular, measuring low energy X-ray signals using a charge couple device (CCD) is challenging, due in part to their reduced sensitivity below 500 eV, and in part to slow image transfer times. Hence, characterizing organic samples by spectro-ptychography at the C 1s (K) edge, O 1s (K edge) or N 1s (K edge) is difficult. Carbon K-edge ptychography was only recently demonstrated for the first time.~\cite{N_Mille_2022} Examples of non-resonant ptychographic imaging of organic materials performed at X-ray energies higher than 500 eV were reported in Ref. ~\cite{M_Rose_2015, K_Giewe_2011, M_Rose_2018,V_Savikhin_2019, M_Beckers_2011, M_Jones_2014}. A recently developed, uncoated, scientific complementary metal-oxide-semiconductor (sCMOS) sensor has excellent performance over the soft X-ray energy range (100 – 2000 eV), high quantum efficiency, low background and fast data transfer (40 frames/s at the full size of the sensor - 2048 $\times$ 2048 pixels). An extensive description of the detector,~\cite{Kevin_2020} and results of proof of principle ptychography experiments at the C 1s edge~\cite{N_Mille_2022} have been reported elsewhere.

Motivated by the availability of a suitable soft X-ray camera and ptychography processing capabilities, a soft X-ray ptychography setup was developed at the STXM endstation of the HERMES beamline at Synchrotron SOLEIL.~\cite{R_Belhou_2015} Ptychography capabilities strongly rely on the properties of the instrument, sample and the employed methodology. For example, a high resolution image can be reconstructed from a sample which scatters (diffracts) partially or fully coherent X-rays over large scattering angles (i.e. out to a large values of the elastic X-ray scattering vector, \textbf{q}, where  \textbf{q} = $\frac{4\pi}{\lambda}$ $\sin$($\Theta$), $\lambda$ = X-ray wavelength, $\Theta$ = scattering angle). Methodology optimization may include: spot size, overlap between measured points, sampling pattern, exposure time, and method of subtracting background signal. Most of these factors are known and their optimization has been addressed elsewhere.~\cite{Edo_2013, N_Mille_2022, D_J_Batey_2014, O_Bunk_2008, X_Huang_2017, Liu_2013, C_Wang_2017,M_Dierolf_2010_2} Additional factors become important when measuring radiation sensitive samples, such as organic polymers and biological samples. In such cases trade-offs are necessary to achieve an optimal ptychography measurement, while at the same time maintaining a sufficiently low dose that the results can be attributed to the original material, and not radiation damage products.~\cite{Wang_2009,Jones_WM_2014} This can be achieved by a combination of reduced exposure time, reduced beam intensity and using a defocused probe beam. However, these approaches may reduce the signal to noise ratio which can potentially reduce the quality and spatial resolution of the reconstruction. Hence, one must optimize these parameters such that an efficient ptychography measurement can be performed within an acceptable dose.

Here we identify different measurement and reconstruction parameters to be taken into consideration to perform efficient and successful soft X-ray ptychography for a variety of samples. We describe different criteria based on the results from soft X-ray ptychography on carbon nanotubes (CNT), boron nitride (BN) nanobamboo structures and permalloy nanowires, the former two scatter relatively less compared to the later. We characterize the chemical, physical and electronic properties of these samples based on the microscopy and spectroscopy data obtained from the ptychography measurements. These samples were chosen to cover the following soft X-ray absorption edges: Boron 1s (K edge, 185-220 eV), Carbon 1s (K edge, 280-315 eV), Nitrogen 1s (K edge, 395-420 eV), and Iron 2p (L$_{2,3}$ edges, 700 – 730 eV). In addition, since the soft X-ray range below 500 eV is of interest for measuring radiation sensitive materials such as polymers and biological samples, we describe the ptychography measurement parameters which affect the radiation dose delivered to the sample, and provide an approach to quantitatively estimate that dose. To the best of our knowledge, the B 1s edge results reported here are the lowest energy spectro-ptychography measurements to date using synchrotron X-rays. These capabilities complement laser based, higher harmonic generation (HHG) ptychography at still lower energies.~\cite{B_Zhang_2015, Loetgering_2022}

\section{Samples}
\subsection{Carbon nanotubes}
Ptychography was measured from two different CNT samples. CNT\#1 consists of bundles of single-walled CNT, prepared by a two-stage laser method.~\cite{C_Kingston_2004} It was the subject of an extensive study for its chemical functionalization using STXM, Raman spectroscopy, thermal analysis and other methods.~\cite{E_Najafi_2010} Sample CNT\#2 is arc discharge multi-walled CNT, purchased from Sigma Aldrich and used without further thermal or chemical treatments. Both samples were dispersed in ethanol~\cite{Najafi_2008} and sonicated for 30 seconds to ensure sufficient dispersity without inducing structural damage to the tubes. The nanotube diameters range from 100--200 nm, with length in the order of 5--10 $\mu$m. They were then drop cast onto a formvar coated grid (CNT\#1, measured December 2020) or a 5 nm thick Si$_3$N$_x$ membrane (CNT\#2, measured June 2021), and heated at 50$^{\circ}$C under vacuum at 10$^{-3}$ mbar for 8--10 hours to remove residual solvent.

\subsection{Boron nitride nanostructures}
The born nitride nanobamboo (BNB) structures were prepared by ball milling of elemental boron and LiO$_2$ in NH$_3$ gas, followed by thermal annealing on a stainless steel substrate.~\cite{L_H_Li_2010} Their diameter is typically 100 nm.~\cite{Kruger_2020} The resultant boron nitride nanobamboo structures were scraped from the substrate directly on to a transmission electron microscopy (TEM) grid covered with lacy carbon.~\cite{X_J_Dai_2011}

\subsection{Permalloy nanorods}
The permalloy nanorods were prepared at room temperature by pulsed electrochemical deposition. Details of the procedure are presented elsewhere.~\cite{Ruiz_2018} The synthesized permalloy nanorods have diameters between 100--150 nm and were dispersed in ethanol (99.5\% vol.). Samples were prepared by drop casting on Si$_3$N$_x$ membranes.

\subsection{Characterization techniques}

Both STXM and ptychography were measured on the same areas in order to investigate the relative merits of these methods. For ptychography the single channel STXM detector was replaced with a soft X-ray sensitive camera (a customized Tucsen Dhyana 95 sCMOS camera). An uncoated sensor was used by the studies below 500 eV, while a coated sensor was used for studies above 500 eV. Further details of the sensors are given elsewhere.~\cite{Kevin_2020} First the sample is imaged using the STXM configuration during which images and multi-energy image sequences (stacks) are recorded to identify the resonant energies and coordinates of the regions of interest. Then the phosphor/PMT detector used for STXM is replaced with the Dhyana camera and a suitable filter.~\cite{N_Mille_2022} The Dhyana camera is operated with 100--200 ms dwell to acquire each diffraction pattern (recently upgraded Dhyana cameras are able to operate  at least 2x faster). For every ptychographic image (which is derived from a set of diffraction images (DI) measured with overlapping beam spots) the dark signal (average of 25 camera images measured without X-rays) is subtracted from each DI. The typical dark signal, averaged over the full camera image, is ~2000 counts/s – about 3\% of the 16-bit dynamic range.  More typically, lower intensities are used to limit radiation dose and exposures are 100 ms. An example of signal and background camera images from a 1.0 mm defocused measurement of BNB is presented in \textbf{Supporting Information (SI)} - section S.1 and Fig. S1. For stack measurements (multi-energy imaging) only one dark image is acquired and used for the whole stack.  For STXM the fully focused beam is used (in this work the full-focus probe is a 62 nm diameter spot generated by a zone plate with outer zone width of 50 nm).  For most of the ptychography measurements, the sample or the zone plate is shifted away from the in-focus position along the X-ray propagation axis in order to measure ptychography using a defocused spot. For the 1 $\mu$m spot used in this work, the illumination is an annulus with outer diameter of 1 $\mu$m and inner diameter of 0.3 $\mu$m at the sample.

The HERMES beamline consists of two APPLE II undulators which provide linear vertical (LV) and linear horizontal (LH) polarized X-rays for X-ray linear dichroism (XLD) studies, and  circular left (CL) and circular right (CR) polarized light for X-ray magnetic circular dichroism (XMCD) studies.  For XLD and XMCD characterization, images with LH and LV (or CR and CL) are acquired separately. After ptychographic reconstruction, the amplitude and phase images are aligned to correct for the drift. The amplitude images are converted to optical density (OD) using the I$_o$ signal from a region of the amplitude images without the sample. The final dichroic images are calculated as the difference between OD images of (LV – LH) for XLD and (CL – CR) for XMCD. XLD provides information on the asymmetric geometry of a sample through its effect on the intensities of specific electronic transitions.~\cite{H_Ade_1993} XMCD provides information on the local magnetization of the ferromagnetic material.~\cite{Stohr_1999} Polarization dependent characterization of the XLD of CNT and BN can be used to identify point defects and local electronic properties of the material, which are of potential interest in electronic device applications.~\cite{Felten_A_2010, Dai_H_2002, Chopra_N_G_1995} XMCD is an important characterization technique in magnetism and magnetic storage devices as it can be used to quantify the magnitude of the local spin and orbital moment~\cite{Shridhar_2018} and measure the magnetism of ferromagnetic domains and domain walls.~\cite{Jaianth_2020, Jaianth_2019} Ptychography reconstruction is performed using the Python tools for Nano-structures Xtallography (PyNX).~\cite{Favre_Nicolin_2020, Mandula_2016_pynx, Favre_Nicolin_2011} PyNX offers several complementary reconstruction algorithms including alternate projection (AP),~\cite{Marchesini_2014_AP} difference map (DM),~\cite{P_Thibault_2009_DM} and maximum likelihood (ML) methods.~\cite{Odstrvcil_LH_2018, P_Thibault_2012_max} Supporting Table S1 summarizes the experimental acquisition and PyNX reconstruction parameters used to obtain the results presented in this work.

\section{Results}
\subsection{Carbon nanotubes}
Since the results for the two carbon nanotube (CNT) samples are similar to those reported earlier~\cite{N_Mille_2022}, we present those results in the \textbf{SI}. Figure S2(a) shows a STXM image of the CNT\#1 sample measured at 350 eV using a 25 nm outer diameter zone plate. Figure S2(b) is the amplitude and Fig. S2(c) is the phase ptychography image, reconstructed from a set of diffraction images (DI) recorded with a 1 $\mu$m spot size at 285.2 eV using linear horizontal (LH) polarized light. The dark square in the upper right corner of the ptychography images is carbon build-up from an earlier ptychography measurement using a focused spot. The contrast in the phase image is enhanced compared to that of the amplitude image. One can distinguish overlapping CNT due to the strong phase contrast at the edges of the CNT.  Figure S2(d,e) show ptychography absorption images of CNT\#2, measured at 285.2 eV with LV and LH polarization respectively. Figure S2(f) is the XLD map, obtained as the difference of the ptychography images shown in Fig. S2(d) and Fig. S2(e). In the (LV – LH) XLD map CNT with horizontal orientation have white contrast while CNT with vertical orientation have dark contrast. The XLD contrast of a CNT is strongest at the C 1s $\rightarrow \pi ^*$ transition at 285.2 eV, which has highest intensity when the X-ray polarization is perpendicular to the long axis of the CNT, consistent with the contrast in Fig. S2(f). There is also considerable XLD contrast at the C 1s $\rightarrow$ $\sigma ^*$ transition at 293 eV, which has highest intensity when the X-ray polarization is parallel to the long axis of the CNT.~\cite{R_E_Medjo_2009, Schiessling_2003} The C 1s absorption spectra of CNT\#2 measured with LH and LV polarization, and an alternate presentation of the XLD mapping of by spectro-ptychography are presented in the \textbf{SI} section SI-3 and Fig. S3.

\subsection{Boron nitride nanobamboo structures}

\begin{figure}
	\includegraphics[width=0.8\linewidth]{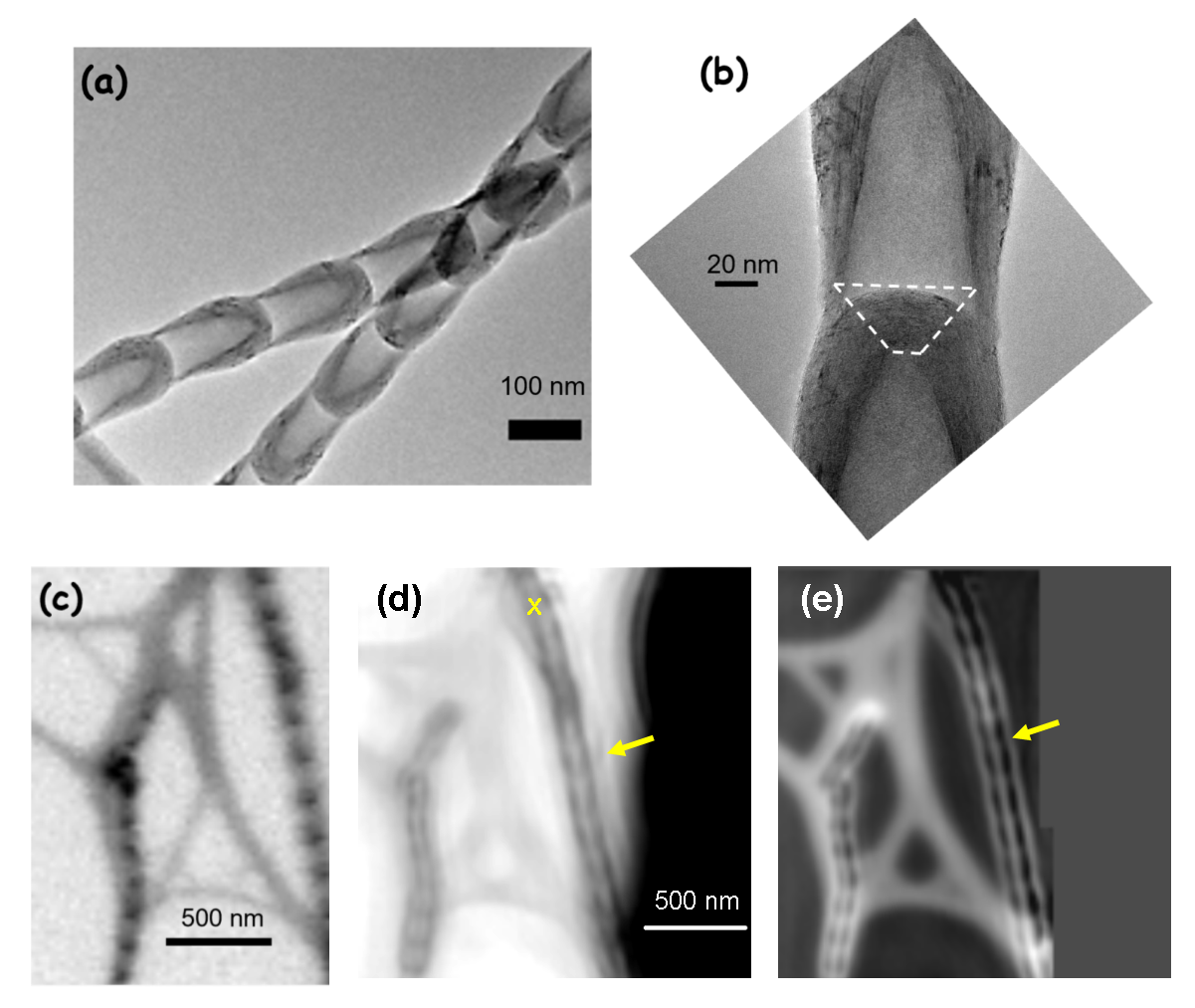}
	\caption{\textbf{Comparison of TEM, STXM and ptychography imaging of BN nanobamboo.} (a) TEM image of the BN nanobamboo structure. (b) Expanded image of the interface between two panes of the bamboo structure. The dotted line indicates the region where the sheets are aligned perpendicular to the long axis. (a, b are courtesy of C. Bittencourt). (c) STXM image of BN nanobamboo structure at 192.0 eV, LH polarization. (d) Ptychography amplitude image of the same BN nanobamboo structure as in (c) (E = 192.0 eV, LH polarization). The yellow ‘X’ marks where the DI presented in Fig. S6 was measured. (e) Phase image of the same area as in (d), derived from the same ptychography data. The yellow arrow in (d, e) highlight the significant differences in the details of the nanobamboo structure between the amplitude and phase image.}\label{fig:fig1}

\end{figure}

The structure of BN nanobamboo (BNB) is more complex than that of CNT. Figure~\ref{fig:fig1}(a) shows a bright field TEM image of the BN nanobamboo structures. The characteristic bamboo structure is a result of the ball milling process where the nano-sheets take a conical shape in each sub-unit/element of the bamboo structure and subsequent sections are formed on top of the conical structure, resulting in a long chain of the bamboo like structure. Figure~\ref{fig:fig1}(b) shows a zoomed-in TEM image of two panes for better visualization of the sheet alignment. The sheets are aligned vertically at the edges but are bent horizontally at the interface between two bamboo elements, as highlighted by the dotted line in Fig.~\ref{fig:fig1}(b).~\cite{Terrones_2007} The thickness of the BN at the horizontal region (marked by the dotted line) in the long axis direction is 25--30 nm, while the thickness on the sides along the long axis is 16--40 nm. The orientation of these nanobamboo structures relative to the X-ray polarization direction gives rise to the XLD effect. The total thickness of the nanobamboo is 80--120 nm.

Figure~\ref{fig:fig1}(c) shows a STXM transmission image of BN nanobamboo measured with the 50 nm zone plate in focus at 192.0 eV with LH polarization. This STXM image was measured using the camera, with a step size of 20 nm in an area of 2 $\times$ 2 $\mu$m$^2$. The camera images are later integrated to form the STXM image. This approach results in an image similar to that measured by STXM using a single point detector. The STXM image shows the nanobamboo structure as a series of black dots along the length of the nanobamboo, and the edges not resolved. Figure~\ref{fig:fig1}(d) shows the reconstructed ptychography amplitude image while Fig.~\ref{fig:fig1}(e) shows the phase image of the same area, from a data set recorded at the B 1s $\rightarrow \pi ^*$ transition at 192.0 eV. It is noteworthy that BN nanobamboo sheets are better resolved in the phase images than the amplitude image [see yellow arrows in Fig.~\ref{fig:fig1}(d) and \ref{fig:fig1}(e)]. In particular, the central low density region is clearly differentiated from the denser border structures in the phase image.

\begin{figure}[tbh]
\includegraphics[width=0.8\linewidth]{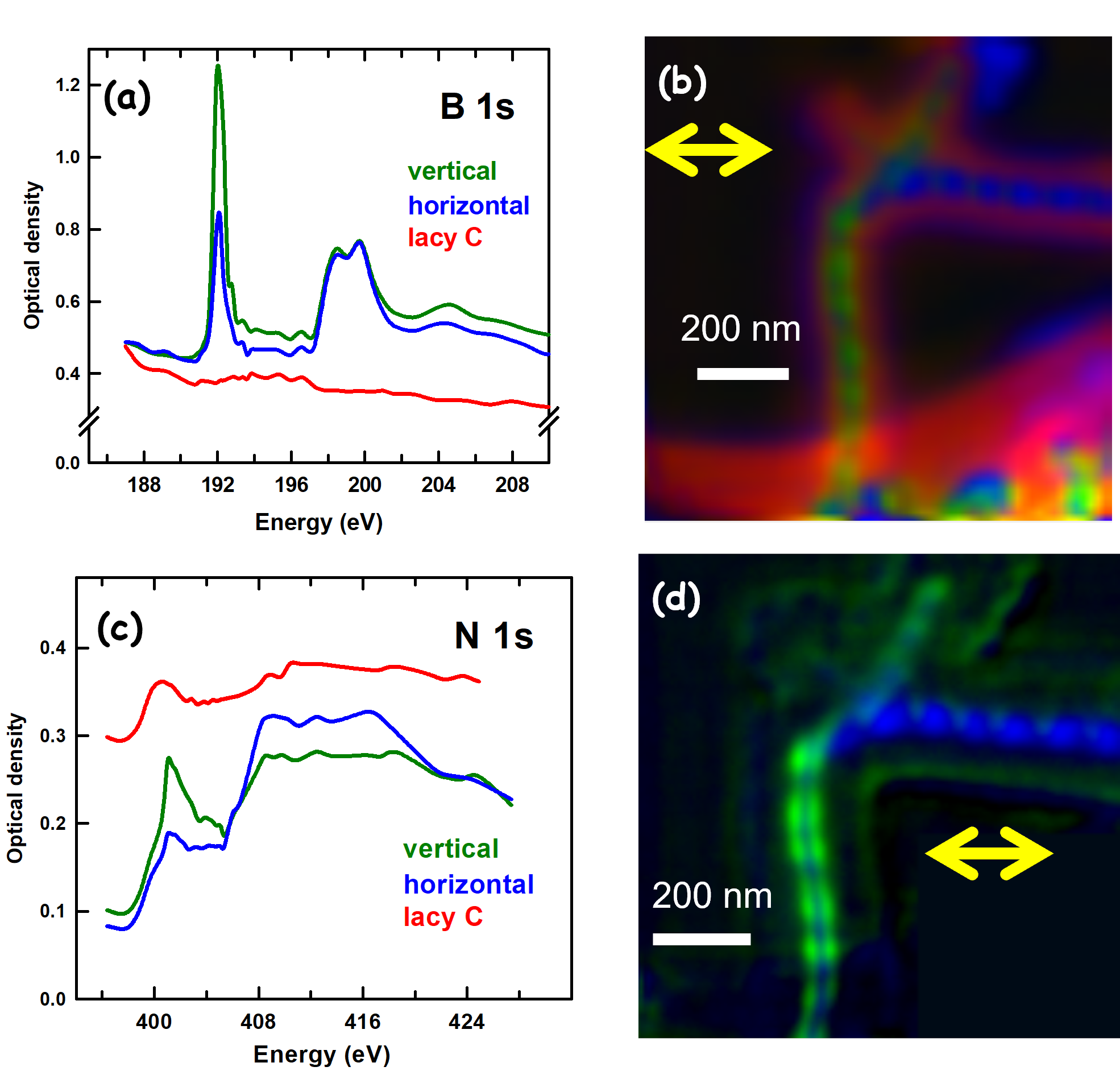}
\caption{\textbf{Spectroscopy and dichroic mapping of the BN nanobamboo sample from spectro-ptychography.} (a) B 1s absorption spectra from amplitude ptychography reconstruction. The spectra of vertical (green), horizontal (blue) BN nanobamboo and the lacy carbon support (red) measured with LH polarized light. (b) Chemical and dichroic mapping from fits of the B 1s stack (42 images from 178 – 210 eV) to the spectra in (a) (same color coding). (c) N 1s absorption spectra from amplitude ptychography reconstruction. The spectra of vertical (green), horizontal (blue) BN nanobamboo and the lacy carbon support (red) measured with LH polarized light. (d) Chemical and dichroic mapping from fits of the N 1s stack (43 images from 396 – 428 eV) to the spectra in (c) (same color coding).}\label{fig:fig2}

\end{figure}

Figure~\ref{fig:fig2}(a) and Fig.~\ref{fig:fig2}(c) shows the B 1s and N 1s absorption spectra of BN nanobamboo, measured by spectro-ptychography. The ptychography data sets were measured as a function of different photon energies, first with LH polarization, then LV polarization. The absorption spectra are extracted from the image pixels of the central regions of vertical and horizontally aligned nanobamboo structures. The spectra presented in Fig.~\ref{fig:fig2}(a) and Fig.\ref{fig:fig2}(c) are extracted only from the stack measured with LH polarization. The green and blue spectra are taken from the pixels of the green and blue regions of the nanobamboos in Fig.~\ref{fig:fig2}(b) and Fig.\ref{fig:fig2}(d), respectively. The red spectrum in Fig.~\ref{fig:fig2}(a) and the red regions in Fig.~\ref{fig:fig2}(b) corresponds to the lacy carbon support. The B 1s $\rightarrow \pi^*$ peak at 192 eV is strong in the spectrum recorded with the E-vector perpendicular to the BNB axis (LV for horizontal BNB, LH for vertical BNB) but has lower intensity in the opposite polarization. The XLD intensity (i.e. the difference between the peak with LH and LV polarization) is $\sim$50\% of the peak intensity.

Figure~\ref{fig:fig2}(c) shows the N 1s absorption spectrum of BNB obtained from spectro-ptychography data using LH polarization. The blue and green lines shown in the spectra are obtained from the regions highlighted in the same color in Fig.~\ref{fig:fig2}(d). The characteristic N 1s $\rightarrow \pi^*$ absorption peak at 401.3 eV shows an XLD signal of $\sim$30\%.~\cite{Fuentes_2003} Furthermore, the XLD signal in the region of the N 1s $\rightarrow \sigma^*$ transitions (408–416 eV) is stronger than the XLD signal in the B 1s $\rightarrow \sigma^*$ transition region (198–202 eV). The relatively weak XLD for the $\sigma^*$ peaks is probably due to the curvature of the bamboo which mixes the $\sigma^*$ and $\pi^*$ levels resulting in higher absorption when X-ray polarization is aligned in the perpendicular direction of the long axis of the BNB. Electronic structure calculations, such as those recently reported for closely related BN nanotube structures,~\cite{Fuentes_2003, Kruger_2020} are required for a deeper understanding of these spectra and the associated XLD. 

\begin{figure}[tbh]
\includegraphics[width=0.8\linewidth]{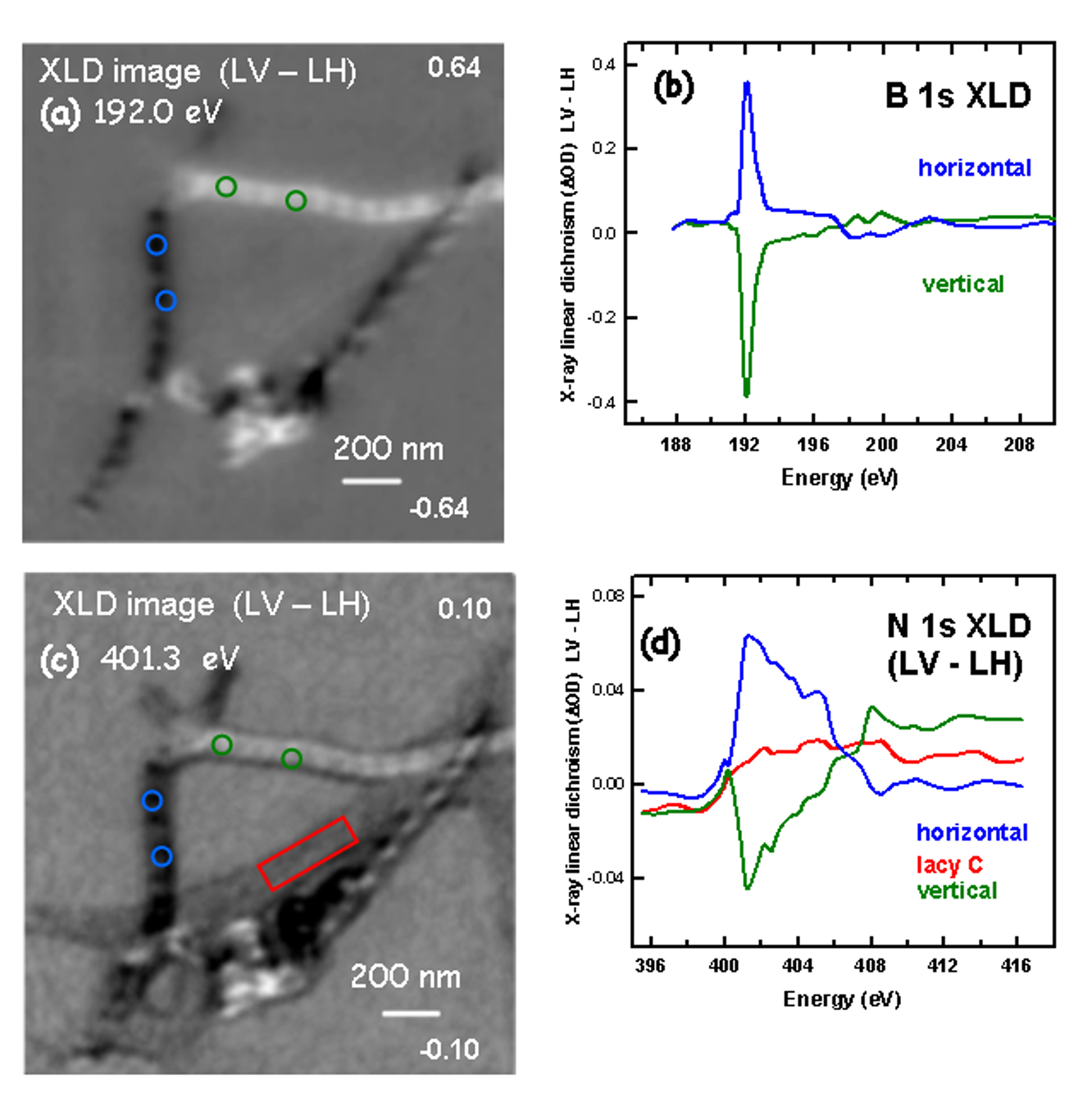}
\caption{\textbf{XLD mapping of BN nanobamboo.}  (a) XLD image (LV – LH) measured at 192.0 eV, the B 1s $\rightarrow \pi ^* $ transition. (b) B 1s XLD spectra for horizontal and vertical BN nanobamboo. (c) XLD image measured at 401.3 eV, the N 1s $\rightarrow \pi ^* $ transition.  (d) N 1s XLD spectra for horizontal and vertical BN nanobamboo. The numbers at the right of each XLD map are the intensity limits ($\Delta$OD).}\label{fig:fig3}

\end{figure}

Figure~\ref{fig:fig3}(a-d) present the X-ray linear dichroism (XLD) results for BNB derived from ptychography at the B 1s and N 1s edges, respectively. Here the XLD map (LV – LH) was obtained from the difference in ptychography absorption images measured with LH and LV polarization. One can find in both Fig.~\ref{fig:fig3}(a) and Fig.~\ref{fig:fig3}(c) alternate bright and dark contrast across the bamboo structure. The horizontal BNB has opposite contrast from that of the vertical BNB. The regular distribution of alternating contrast suggests a uniform size distribution of the nanobamboo sections during the growth process. From the XLD image the individual panes can be identified; these are not as visible in the individual ptychography amplitude images. The XLD contrast appears to originate from the individual panes as single units, which are determined by the orientation of the whole bamboo structure. As shown in Fig. S4, there are rapid changes in contrast in the BN nanobamboo structures as the photon energy is scanned across the B 1s $\rightarrow \pi^*$ (and N 1s $\rightarrow \pi^*$, not shown) peaks. In addition, there is a shoulder on the high energy side of each $\pi^*$ peak, which has different XLD properties from that of the main peak. The origin of these spectral details requires additional studies using quantum mechanical calculations like those reported for BN nanotubes,~\cite{Fuentes_2003, Kruger_2020} and will not be discussed further here.

\subsection{Permalloy nanorods}
\begin{figure}[tbh]
\includegraphics[width=0.7\linewidth]{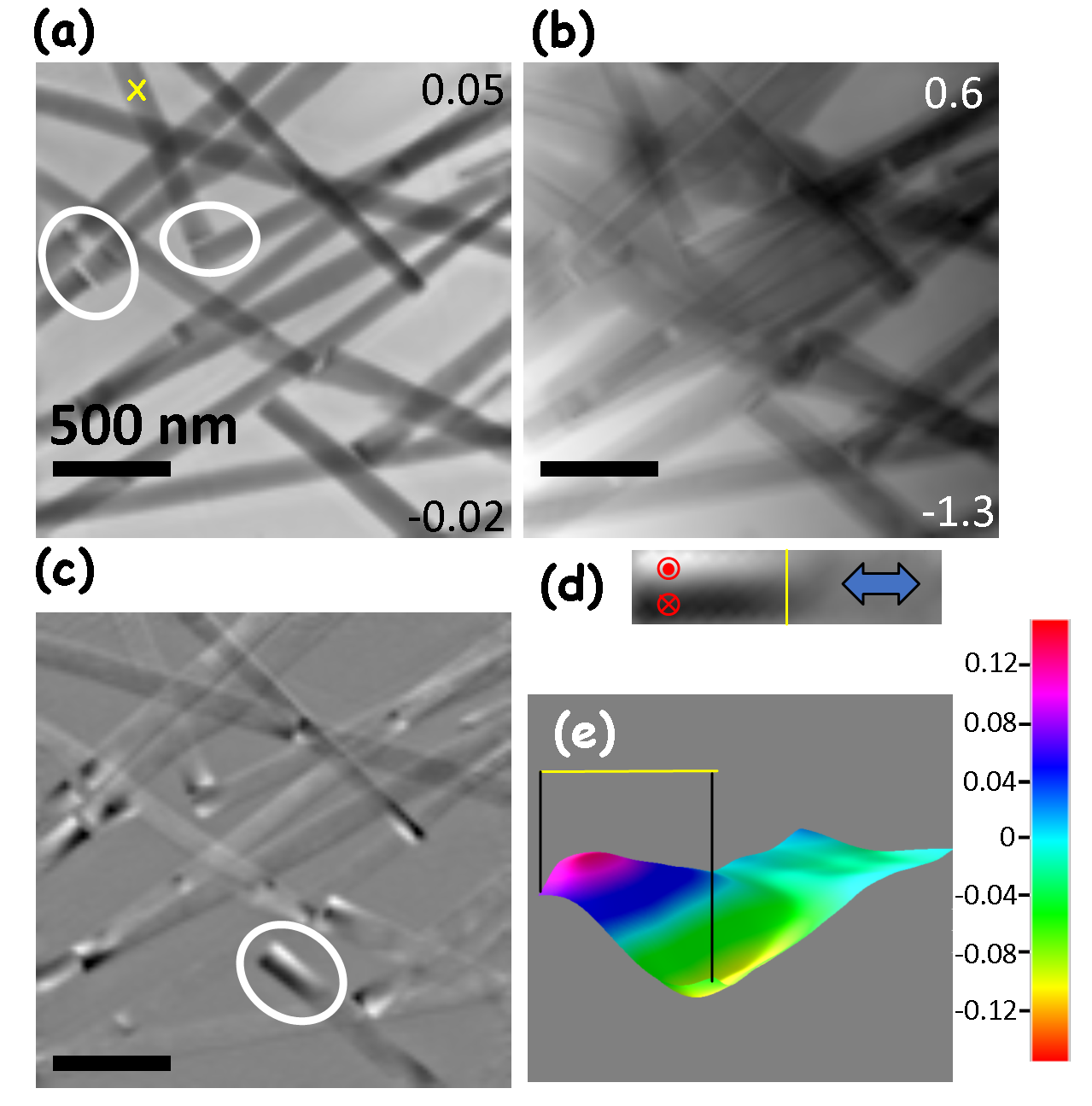}
\caption{\textbf{Spectro-ptychography and XMCD of permalloy nanorods.} (a) Ptychography amplitude image of permalloy nanorods measured at 706 eV using CL polarized X-rays. The white circles indicate gaps between the nanorods in the range of 20 – 30 nm.  The yellow ‘X’ marks the region where the DI presented in Fig. S6 was measured. (b) Ptychography phase image of the same area as in (a). (c) XMCD image of the nanorods with contrast ranging from $\pm$0.15. The scale bar in (a-c) corresponds to 500 nm. (d) Magnified image of the termination of the nanorod, located in the white circle in (c). (e) Surface plot of (d) with color scale indicating the local XMCD at the termination of the nanorod. The yellow lines in (d) and (e) correspond to the width of the rod which is 140 nm.}\label{fig:fig4}

\end{figure}

Magnetic materials with 3D structures such as permalloy nanorods are of potential interest for their capability to stabilize topological magnetic spin structures. Figure~\ref{fig:fig4}(a) shows the reconstructed ptychography amplitude image of permalloy (Ni$_{81}$Fe$_{19}$) nanorods acquired with CL polarization, at a photon energy of 706 eV (Fe 2p$_{3/2}$ peak). Figure~\ref{fig:fig4}(b) is the reconstructed ptychography phase image. From the amplitude image one can observe spacings of 20--30 nm between adjacent nanorods [examples indicated by the circles in Fig.~\ref{fig:fig4}(a)]. In order to estimate the spatial resolution, a line profile over the edge of the nanorod in Fig.~\ref{fig:fig4}(a) is presented in Fig. S5. From the 20\%--80\% jump we estimate the resolution to be 16 nm. In the phase image [Fig~\ref{fig:fig4}(b)], there is a dark contrast region at the sides of some of the nanorods. The origin of these additional contrast features is unclear. Nevertheless, in the phase image, the ends and sides of the nanorods are sharp and have a darker contrast than the central region. Enhanced contrast at the edges/sides can also be seen in the amplitude image, but less clearly than in the phase image. 

Magnetic structures such as the permalloy nanorod typically have a shape anisotropy which aligns the magnetization towards the long axis of the rod, minimizing the anisotropy energy in the system. However, the ends of the nanorods are terminated by sharp edges which can result in a multi-domain state and can induce formation of topological spin structures.~\cite{Fert_2017,X_Shi_rev_2019,Ruiz_2020} Figure~\ref{fig:fig4}(c) shows the ptychographic amplitude XMCD image of the nanorods calculated as the difference between the OD images acquired with CR and CL polarization, and normalized to the sum of the images. The XMCD contrast is related to the magnitude and orientation of the local magnetization vector with respect to the photon spin vector, which lies along the X-ray propagation direction. XMCD intensities vary as a cosine function [$\approx$ \textbf{M} $\cdot$ \textbf{P} $\cdot$ $\cos\theta$, where $\theta$ is the angle between the magnitude of the photon spin (X-ray propagation) vector (\textbf{P}) and the magnetization vector (\textbf{M})]. Since the nanorods are orthogonal to the X-ray beam, the edges (circular termination of the nanorods) with possible topological spin structures are not visible. Nevertheless, close to the termination of the nanorod [highlighted by a white circle in Fig.~\ref{fig:fig4}(c)], the uniform grey contrast splits into white and dark contrast, indicating a change in the magnetization orientation near the termination. Figure~\ref{fig:fig4}(d) is a magnified image of the region in Fig.~\ref{fig:fig4}(c) encircled by the white line. Here the grey contrast indicates the magnetization is in the plane of the substrate along the long axis of the rod, while the white and dark contrast indicate magnetization pointing parallel and anti-parallel to the direction of the X-ray propagating vector. The dark and light contrast corresponds to an XMCD signal of $\pm$15\%. Typically, if the magnetization of Fe aligns parallel to the X-ray propagation vector (with CR or CL polarization) one can expect an XMCD signal of $\pm$30--35\%. However, the observed XMCD intensity of the nanorods is considerably weaker. It appears that the magnetization is still in-plane with respect to the nanorod, with some tilt. The color scale image shown in Fig.~\ref{fig:fig4}(e) represents the XMCD signal in Fig.~\ref{fig:fig4}(d) as a surface plot. This representation shows that there is a smooth transition (white – grey – black) where the magnetic domain is split at the termination.

\section{Discussion}

\subsection{Controlling extent of radiation damage in low energy ptychography}
Carbon contamination and radiation damage is an important issue in ptychography. To reduce the total exposure time on the sample and acquisition time one can use a defocus beam.~\cite{Rodenburg_2019, Song_2019} Even though defocus ptychography is well known and widely used, for low scattering sample there are a few challenges to overcome. In the next section we illustrate the effect of defocus beam on the radiation dose, describe the dose associated with  beam overlap, and show the effect of using a defocused beam on the quality of the reconstructed images.

\subsection{Beamsize and overlap}
\begin{figure}[tbh]
\includegraphics[width=\linewidth]{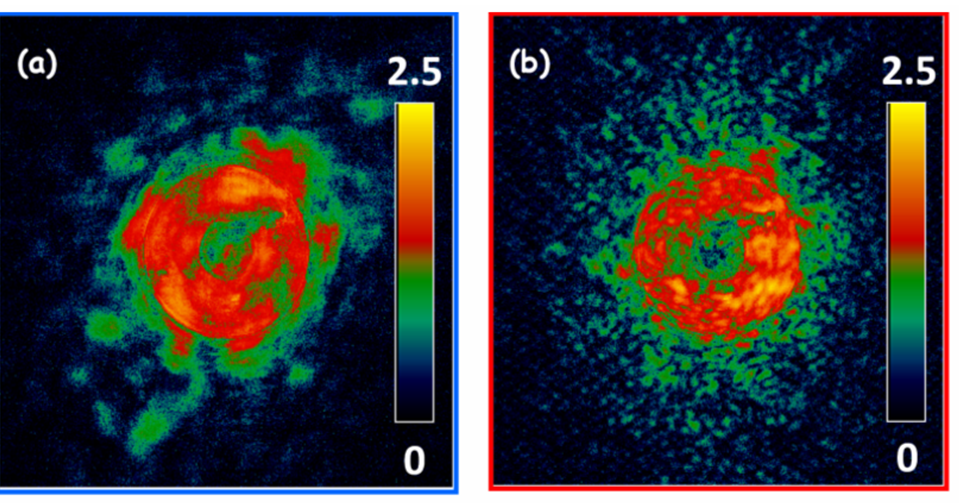}
\caption{\textbf{Effect of spot size on diffraction images(DIs).} (a) X-ray scattering signal from BN sample using focused beam (diameter of 62 nm).  (b) X-ray scattering signal from the same spot as (a) using defocused beam of diameter 1000 nm. Both DIs were recorded at 192 eV and LH polarization, and a sample-camera distance of 51.3 mm.}\label{fig:fig5}

\end{figure}

Starting from focused conditions, the size of the beam is adjusted by moving the sample upstream or the Fresnel zone plate downstream along the direction of X-ray propagation. The displacements is typically less than 50 $\mu$m. Figure~\ref{fig:fig5} shows how the DI changes when the same spot on the BN sample is illuminated using focused (62 nm) and defocused (1000 nm) beam. The scattering signal measured using a defocused beam [Fig.~\ref{fig:fig5}(b)] extends to larger \textbf{q} (farther from the centre) than that recorded using focused beam [Fig.~\ref{fig:fig5}(a)] We attribute this to scattering from regions outside the area exposed by the focused beam. Such extended signals, combined with appropriate overlap between adjacent diffraction signals, are typically considered beneficial in ptychography reconstruction and can potentially enhance the spatial resolution.~\cite{O_Bunk_2008}  

\begin{figure}
\includegraphics[width=0.8\linewidth]{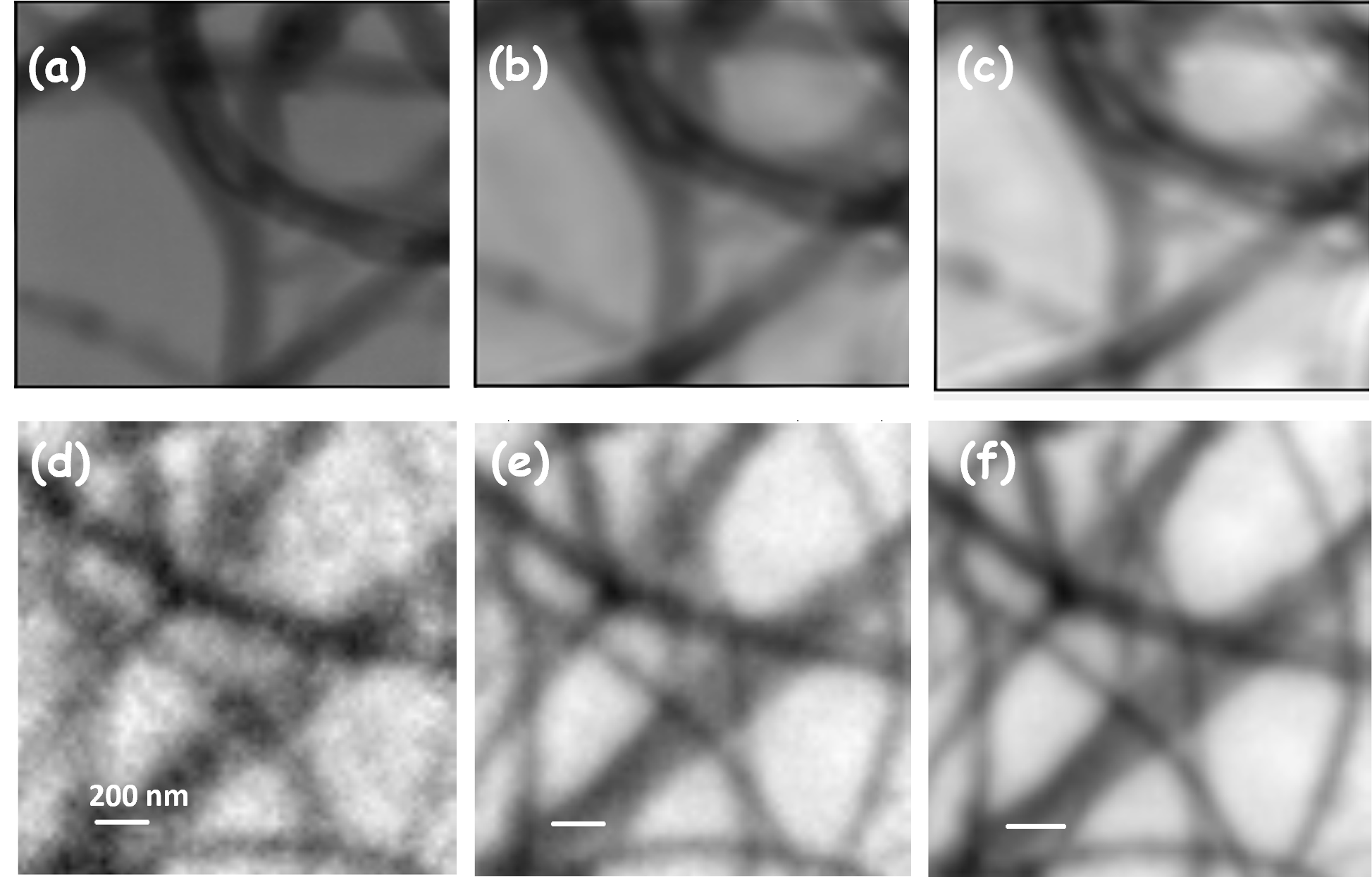}
\caption{\textbf{Effect of beam size and overlap on reconstructed images.} Reconstructed CNT\#2 amplitude image measured at 310 eV (LH polarization, in an area of 1.6 $\times$ 1.6 $\mu$m$^2$) measured using (a) focused beam – 62 nm (67 \% overlap), (b) defocused beam - 500 nm (92 \% overlap) and (c) defocused beam - 1000 nm (98\% overlap) with detector positioned at 56 mm. The sharpness measured from the variance of the Laplacian transform is reduced by about 14\% in (b) and (c) with respect to (a). Reconstructed CNT amplitude image measured in an area of 1.6 $\times$ 1.6 $\mu$m$^2$ at 285.2 eV (LH polarization) and a 500 nm diameter defocused beam, recorded with (e) 10 $\times$ 10 points (68 \% overlap); (f)  20$\times$20 points (84 \% overlap), and (g)  40$\times$40 points (92 \% overlap).Details of the acquisition and reconstruction parameters are presented in table S-1.}\label{fig:fig6}

\end{figure}

Figure~\ref{fig:fig6}(a-c) shows the difference between the quality of the reconstructed images of CNT using focused and defocused beam. The sharpness of each image was measured by the Laplacian operation~\cite{Git_page} which typically measures the gradient at edges. The variance of the Laplacian is related to the sharpness of the image, with lower values indicating higher sharpness. Compared to the focused image [Fig.~\ref{fig:fig6}(a)] the sharpness is reduced by about 14\% for both the 500 nm [Fig.~\ref{fig:fig6}(b)] and 1000 nm [Fig.~\ref{fig:fig6}(c)] defocused beam. Subtle differences can still be seen between the 500 nm and 1000 nm images with the former showing some clear features on the CNT structures. With a larger amount of sample being exposed to the defocused beam, it is possible to increase the step size to further reduce the delivered radiation dose on the sample. However, for weakly scattering samples as CNT or BN, we observe that an overlap of at least 90\% is needed to have an effective reconstruction. This is demonstrated in Fig.~\ref{fig:fig6}(d-f) which compare reconstructed amplitude images of the same area using a 500 nm beam size and three different overlap values. The image quality (both spatial resolution and contrast) improves with increasing overlap. Laplacian evaluation shows a sharpness increase of about 15\% when the overlap is increased from 68\% to 92\%. We determined the signal-to-noise ration (SNR) of the CNT signal from the ratio of the OD of the CNT and its standard deviation in the image. We find that the SNR increases by 21\% between 68\% and 84\% overlap. The change in SNR between 84\% and above 92\% is very subtle with a maximum change of 0.1\%. While we see large changes in the SNR, the contrast difference, i.e. the ratio of CNT and background, is similar for regions of thick CNT. We conclude that, when using a defocused beam, a high degree of overlap is required for low scattering samples. However, for permalloy nanorods a focused beam size of 62 nm and an overlap of 30\% (results not shown) is already enough to obtain images with good SNR and contrast, similar to the results shown in Ref.~\cite{N_Mille_2022} for ptychography of a Siemens star test sample measured at 280 eV. Nevertheless, we find that for low scattering samples, when using a defocus beam, a high level of overlap (90\% and more) is needed for good reconstruction quality. Since a high degree of overlap increases the radiation dose, an analysis of how the radiation dose depends on beam size and overlap is presented in the next section.

\subsection{Radiation dose}

\begin{figure}[h!]
\includegraphics[width=0.8\linewidth]{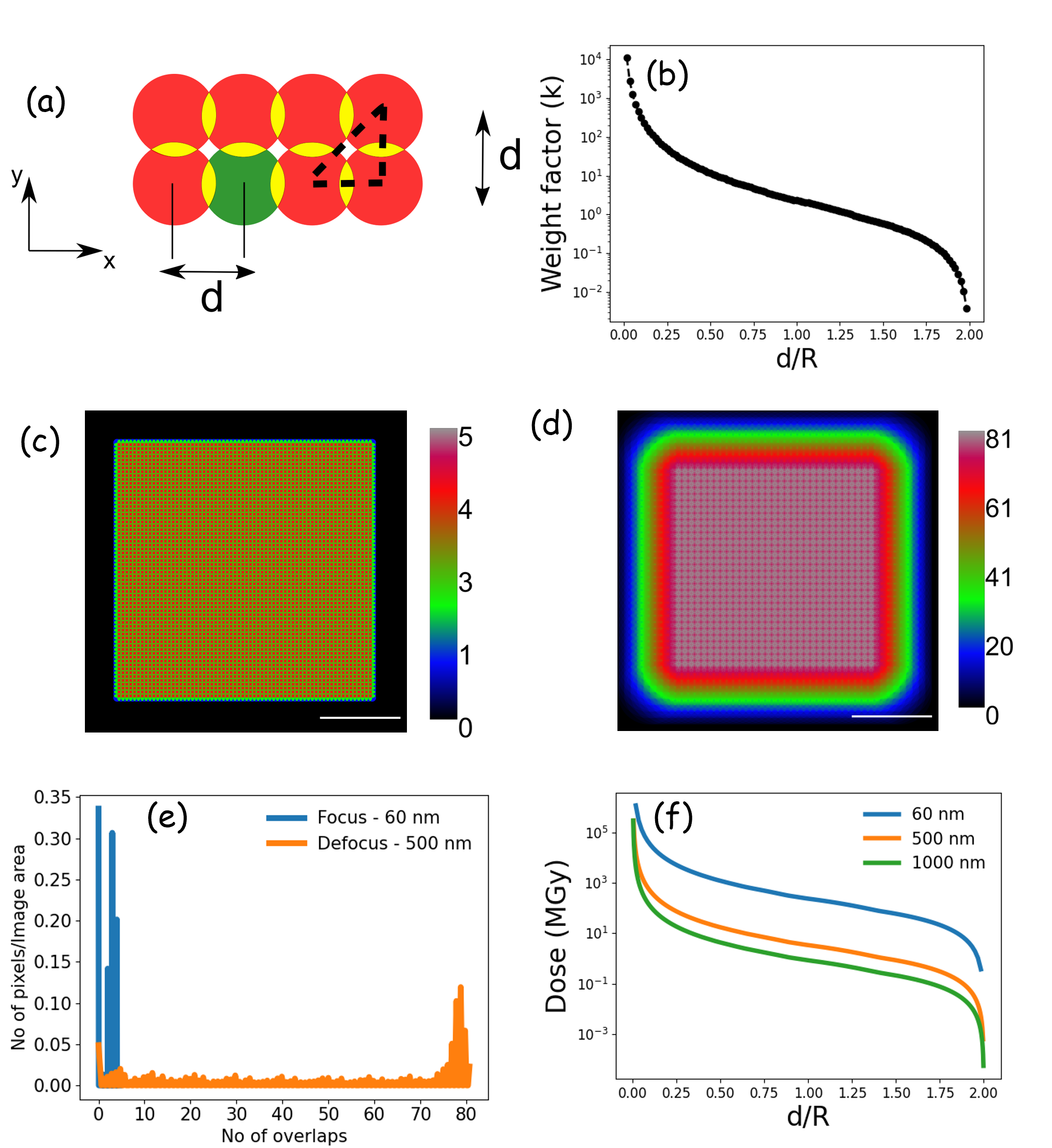}
\caption{\textbf{Estimation of the radiation dose delivered in a ptychography measurement.} (a) Schematic of overlap of X-ray spot, yellow color represents the region of overlap, green is indicated to identify a single X-ray beam spot. (b) Plot of weight factor as a function of d/R. (c) Schematic of overlap regions of a focused beam of 60 nm diameter with step size of 30 nm in a 2 $\times$ 2 $\mu$m$^2$ region. The color bar represents the number of times a region is exposed. (d) Schematic of the overlap regions of defocused beam of 500 nm diameter with step size of 50 nm in a 2 $\times$ 2 $\mu$m$^2$ region. The color bar represents the number of times a region is exposed. The scale bar in (c) and (d) represents 500 nm. (e) Histogram of the number of overlap of a focused and defocused beam shown in (c) and (d). (f) Plot of dose change on BN sample at 192 eV using different beam sizes.}\label{fig:fig7}

\end{figure}

Here we develop an analytical approach to quantify the radiation dose as a function of overlap through simple geometric considerations. We consider overlap geometries as a result of raster scanning, however, the same approach can also be applied to estimate the overlap from non-raster scans. The overlap ratio (O) is typically defined by~\cite{O_Bunk_2008, X_Huang_2017}: 

\begin{equation}
	O = 1 - d/2r	
\end{equation}

where $d$ is the centre-to-centre distance of adjacent spots and 2$r$ is the diameter of the spot. The overlap value is crucial for ptychographic reconstruction. Typically, increasing the overlap results in faster convergence, improved contrast, and higher spatial resolution in the reconstructed image.~\cite{C_Wang_2017,P_Thibault_2008,Eric_W} To quantitatively determine the area of overlap we consider the expression of the area associated with two overlapping circles whose schematic is shown in Fig.~\ref{fig:fig7}(a), where the red dots represent the X-ray beam spot on an evenly spaced region $d$. The area of the overlapping region (with circles of same radius) shown in yellow can be expressed as \cite{Eric_W}:
\begin{equation}
    A_{Overlap} = 2R^{2}\cos^{-1} \frac{d}{2R} - \frac{d}{2}\sqrt{4R^2 - d^2}
\end{equation}

The overlap with the neighbouring circles occurs horizontally, vertically and diagonally. The number of circles ($N_{H/V}$) overlapping a single circle [eg. green spot in Fig.~\ref{fig:fig7}(a)] with a given $d$ in one direction (eg. +x and (+x,+y) diagonal direction) is given by;
\begin{equation}
N_{H/V} = 2R/d
\label{eqn:NH} 
\end{equation}

When the number of circles overlapping a given point is known, the total overlap is given by;
\begin{equation}
A = \sum_{i,j=-N_{H/V}}^{i,j=N_{H/V}} 2R^{2}\cos^{-1} \frac{d_{i,j}}{2R} - \frac{d_{i,j}}{2}\sqrt{4R^2 - d_{i,j}^2 }
\end{equation}

where $d_{i,j}$ is the magnitude of the position vector in $x$ and $y$ coordinate (i.e. $d^{2}_{i,j}$ = $x^2$ + $y^2$ ). If $d_{i,j}$ $>$ 2$r$ or $d_{i,j}$ $>$ (2$r$)$^{1/2}$ in the diagonal direction, then the overlap will be undefined due to the $\cos^{-1}$ term which lies between $\pm$ 1. The summation is kept between $\pm$ N$_{H/V}$ to determine the area from the four quadrants. Once the area of the overlapping region is determined, the increased dose per one exposed region is related to an overlapping/weight factor ($k$) which can be expressed as:
\begin{equation}
k = \frac{A}{\pi\cdot R^2} 
\end{equation}

Figure~\ref{fig:fig7}(b) shows the change in $k$ (plotted on a log scale as a function of $d/R$ ($d$= step size, $R$ = radius of the beam). If $d<<R$ (e.g. $d$ = 0.2$R$), the weight factor can be as high as 78 when overlap is about 90\%, suggesting that the dose on a given spot can be 78 times higher than the dose delivered by a single beam without any overlap. As d increases the weight factor decreases until $d$ = 2$R$ where there is no overlap.

In Fig.~\ref{fig:fig7}(c-d) we model the exposure of the beam using a circular focus and defocus beam on a 2 $\times$ 2 $\mu$m$^2$ area. This dimension is used during measurements for BN, for example. In Fig.~\ref{fig:fig7}(c) we consider a beam diameter of 60 nm with step size of 30 nm, which corresponds to 50\% overlap. Here, the regions are exposed between 1–5 times on average during the measurement and the equivalent weight factor of 2.2. Similarly, Fig.~\ref{fig:fig7}(d) describes the case of a defocus beam with 500 nm diameter with step size of 50 nm, corresponding to 90\% overlap. On average the regions are exposed as high as 81 times, and this corresponds to a k factor of 78. In Fig.~\ref{fig:fig7}(c-d) one can find that the number of overlap increase from corner to the middle of the object. Figure~\ref{fig:fig7}(e) shows the histogram of the number of overlaps between Fig.~\ref{fig:fig7}(c) and Fig.~\ref{fig:fig7}(d). With the full focused beam, 30-35\% of the sample area is being exposed 3-5 times. However, with a defocused beam on average 5-10\% of the area is exposed between 70–81 times.

Fig.~\ref{fig:fig7}(b) plots the overlap factor calculated with dimension of the region as $d\cdot N_{H/V}$ $\times$ $d\cdot N_{H/V}$. To calculate the weight factor for the experimental condition we restrict the analysis within the experimental dimension of 2 $\times$ 2 $\mu$m2 to calculate the weight factor and the corresponding dose value. Figure~\ref{fig:fig7}(f) shows the calculated dose as a function of $d/R$ for the 60 nm, 500 nm, 1000 nm diameter beams used in this work. The dose is calculated based on an approach outlined in the supporting material of \cite{N_Mille_2022} using the optical density value from the BN ptychography amplitude data measured at 192 eV in Fig.~\ref{fig:fig1}(b).  The dose delivered by a focused beam and 30\% overlap is similar to that delivered by a defocused beam with 90\% overlap. From this analysis and the preceding section, it is clear that while increasing overlap improves the quality of reconstruction, it partially reduces the benefits of the defocus on reducing the radiation dose. Hence the minimum overlap required for a successful reconstruction needs to be pre-determined for radiation sensitive samples and the acquisition strategy designed to stay within that limit. The factor (‘additional dose’ column) by which overlap increases the dose on the CNT, BN and nanorod samples is listed in Table~\ref{table:ptycho_summary}. Despite having a high $k$-factor (Eqn. 5), when compared with that for focused beam, the total dose is 1 - 2 orders lower with defocused beam and the same overlap. Therefore, with highly scattering samples using defocus beam can significantly reduce the delivered radiation dose. However, this is not the case for low scattering samples, because higher overlap and/or longer dwell times are needed to get adequate reconstructions. Additionally, with focused beam, even with an overlap of only 30–40\%, there is a large amount of carbon deposition, which is evident when measuring at the C 1s edge (see Fig. S2 and Fig. S6 of \cite{N_Mille_2022}). However, with defocused beam there is no carbon build-up on the samples even after long exposure with 90\% overlap during spectro-ptychography measurements. This clearly indicates that there is a net difference in the flux per unit area (fluence) related with different beam size. Using a defocus beam reduces the radiation damage to the sample.


\section{Reconstruction algorithms: optimizing image quality}
Table~\ref{table:ptycho_summary} summarizes the maximum scattering vector (\textbf{q$_{max}$}) sampled for each measurement. The achieved experimental spatial resolution is partly determined by the scattering ability of a given sample, under specific conditions i.e. the photon energy, polarization and the local structural properties of the sample. The relationship of the maximum scattering angle and the spatial resolution can be approximated by $\lambda/ \sin \Theta$,~\cite{K_Shimomura_2015} where $\lambda$ is the wavelength used and $\Theta$ is the maximum scattering angle. This approximation agrees with the experimental resolution which is determined from the knife edge profile [e.g. Figure S5 in the \textbf{SI}]. For high scattering permalloy nanorods, we find the spatial resolution is two times higher than for the low scattering CNT or BN., which we attribute to the strong scattering of the nanorods at the Fe 2p$_{3/2}$ absorption peak.~\cite{M_Sicarios_2012} For permalloy nanorods, with 62 nm beam and using 50\% overlap, the spatial resolution in amplitude images derived from ptychography reconstruction is  $\sim$16 nm (see Fig. S5). However, for CNT measured using the 62 nm X-ray beam and 92\% overlap, the ptychography image has a resolution of $\sim$37 nm [Fig.6(f)]. The main difference between the two samples is the scattering capabilities. Therefore, to enhance the resolution in ptychography one should consider acquiring additional signal by (i) increasing the exposure time, (ii) increasing the beam intensity, and (iii) optimizing thresholding to isolate true signal from the noise still remaining after background subtraction (see Fig. S1). However, factors (i) and (ii) will increase radiation damage and carbon buildup on the sample. Another interesting approach that could be explored is to use a tilting stage to increase the effective scattering angle ($\lambda/ \sin \Theta$ dependence). With a tilting stage, higher frequency signals can be measured by tilting the sample in both axes.~\cite{K_Shimomura_2015} Alternatively, a small grazing incident angle could also be used to acquire additional scattering signals from each point. These alternatives are yet to be explored.

\begin{table}[h!]

	\centering
	\resizebox{\textwidth}{!}{
		\begin{tabular}{p{2cm} p{1.5cm} p{1.5cm} p{2cm} p{2cm} p{2.5cm} p{2cm} p{2cm} p{1cm} p{1.5cm}}
			\hline
			Element--energy          (eV) &  \textbf{q$_{max}$}   (nm$^{-1}$) & Pixels ($@$) & Expected resolution ($\lambda$/$\sin{\Theta}$ nm) & Detector position (mm) & Experimental resolution$^*$ (nm) & Minimum overlap (\%) & Additional dosage$^{\#}$ (factor) & Beam &Algorithm  \\
			\hline
			\hline
			\\
			Fe-706 & $>$ 0.71 & 1024$^2$ & 17 & 61 & 8 $\pm$ 2  & 30--50 & 0.4--2.3 & F & DM, AP \\
			\\
			N-401.8 & 0.26 & 570$^2$ & 47 & 51.3 & 44 $\pm$ 3  & 90 & 78 & DF & AP \\
			\\
			C-285.2 & 0.33 & 1104$^2$ & 37 & 55 & 30 $\pm$ 3 & 80--90 & 18--78 & DF & AP\\
			\\
			B-192 & 0.30& 1400$^2$ & 40 & 51.3 & 45 $\pm$ 6 & 90 & 78 & DF & AP \\
			\\

			\hline
			\\
		\end{tabular}

	}
\caption{Summary of the maximum scattering vector, expected resolution, detector distance, experimental resolution, minimum overlap required, additional dose (weight factor) due to overlap, beam and algorithm used for ptychography measurements of permalloy nanorods [Fig. \ref{fig:fig4}(a)], CNT [Fig.S2(b)] and BN [Fig.\ref{fig:fig1}(d)]. \\$@$   Range of camera pixels used for reconstruction (square, centered on the centre of the annulus) \\$*$ The experimental resolution is calculated from the knife edge profile – see Fig. S5. \\ \# See Table S1 for the estimated dose delivered for each measurement reported in this work
}\label{table:ptycho_summary}	
	
\end{table}

The ptychography data was reconstructed using both difference map (DM) and alternate projection (AP) algorithms in PyNX. DM and AP, differ in the way the object and probe is updated, AP is faster than DM in terms of time/cycle, however, if it works,  DM provides an earlier convergence. In AP the update is carried out by steep descent method. More details about the algorithm can be found elsewhere.~\cite{Marchesini_2014_AP, P_Thibault_2009_DM} For weakly scattering samples such as BN or CNT, reasonable quality reconstruction of the ptychography data was only possible with AP. With DM, the reconstruction becomes unstable after a few iterations. However, for strongly scattering samples such as the permalloy nanorods, successful and stable reconstructions were achieved with DM. It is unclear why DM did not work for the BN and CNT samples. The differences between the ptychography measurements of the permalloy and BN/CNT samples include: size of beam (focused versus defocused), photon energy, and maximum scattering angle (\textbf{q$_{max}$}). An example of the scattering signal from a single point on the permalloy nanorod and BNB structure is shown in Fig. S6. For the permalloy nanorod (Fig. S6a, S6b) the coherent scattering signal at 706 eV extends to 2/3 of the way to the edge of the camera image  (\textbf{q$_{max}$} = $>$0.71 nm$^{-1}$). For the BNB structure [Fig. \ref{fig:fig1}(a)] the coherent scattering signal at 192.0 eV extends to the boundary of the camera image (\textbf{q$_{max}$} = 0.30 nm$^{-1}$)  In addition, the scattering signal only extends in a particular direction, which is related to the orientation of the BNB structure relative to the E-vector, due to the strong linear dichroism at this photon energy (see Fig.~\ref{fig:fig2}). Although the \textbf{q}-range of scattering is similar in the permalloy and BNB samples, the intensity of the diffraction signal from the permalloy is $\sim$4 times larger than that from BNB. The lower scattering signal in the BNB (and CNT) samples may be the reason for the early instability in the DM reconstruction. The object/probe update in DM has a strong dependence on the Fourier constraint imposed. It is possible that large differences between the simulated diffraction and that of the measured data causes the instability. This issue is typically overcome when object/probe is updated by an error reduction/steepest descent approach such as in AP.~\cite{P_Thibault_2009_DM}

\section{Conclusions}
Ptychographic reconstruction of sets of diffraction images measured below 200 eV using synchrotron X-rays has been achieved for the first time. We have studied CNT, BN and permalloy nanorods by soft X-ray spectro-ptychography, reporting high quality polarization dependent X-ray absorption spectra, as well as XLD and XMCD spectra and maps. We carried out ptychography in the low energy X-ray regime and explored how different parameters such as beam size, overlap factor, and choice of algorithm, affect the outcome. These effects are illustrated with results from weakly scattering BN and CNT samples and compared with results from strongly scattering permalloy nanorods. Typically, in STXM, high resolution images can be obtained by using smaller beam size, small scanning steps and longer dwell time. However, for an effective usage of ptychography it is important that one obtains a reconstructed image with a spatial resolution higher than that which can be achieved with STXM, without a smaller beam size, more data points, or long exposure times. Particularly for samples with low scattering signals, it is challenging to obtain high spatial resolution without compromising one or more of these parameters. Nevertheless, the possibility to reduce the radiation dose on the sample and obtain phase information gives ptychography added value as compared to conventional STXM. For example, the edges and other fine details of the BN nanobamboo structures were better resolved in phase than in amplitude images. For permalloy nanorods, due to large scattering from the sample magnetic domain structures of 10--15 nm in size can be identified in the XMCD images. The dose delivered by a defocused beam, and thus the consequences of excessive dose such as carbon build-up, is less than that delivered by a focused beam. While the dose can be reduced using defocused beam, a high level of overlap is required for weakly scattering samples, which can eventually deliver a dose comparable to that delivered with a focused beam with low overlap. For low scattering samples, the AP reconstruction algorithm, where the object and probe are updated by an error reduction/steepest descent approach, produces a stable reconstruction as it is less dependent on the Fourier constraint in the DM algorithm. In conclusion, while ptychography clearly enhances spatial resolution in samples with strong scattering, for samples with weak scattering, the improvement in spatial resolution provided by ptychography is less dramatic. Methods to efficiently acquire better quality high angle scattering signal should be explored for an effective use of soft X-ray ptychography.









\section*{Acknowledgements}
Research was carried out at the HERMES beamline of Synchrotron SOLEIL. We acknowledge SOLEIL for providing development beamtime for this research. We thank the staff of SOLEIL for their support, in particular Arafat Noureddine for his help with the camera control. We especially thank A. Besson and J.-B. Boy for their technical support, E. Fahri and F.-E. Picca for the computing support, and K. Medjoubi for his help with the reconstruction and fruitful discussions. This work was partially funded by project EU-H2020 of the Nanoscience Foundries and Fine Analysis (NFFA) program. Participation of the Canadian group was supported by the Natural Science and Engineering Research Council of Canada and funding from the Faculty of Science of McMaster University. The authors wish to acknowledge Laura Alvara Gomez, Lucas Perez from Department of Fisica de Materiales, Universidad Complutense de Madrid, Madrid Spain and IMDEA Nanosciencia, Madrid, Spain; Oliver Fruchart from Spintec (University Grenoble Alpes, CNRS, CEA), Grenoble, France for providing the permalloy nanorods. The authors also acknowledge and thank the group of Simard (NRC) for providing the CNT\#1 sample and Dr. Carla Bittencourt for providing the BN nanobamboo sample.


\section*{Additional information}
\textbf{Supporting Information} accompanies this paper at [WEB ADDRESS]
\textbf{Competing interests}: The authors declare no competing interests. 

\section*{Data availability}
All processed data used to generate figures in the main paper and the supporting information are available from the authors on request. The as-recorded raw data for ptychography is very large ($\sim$3 Tb for this study) and is deposited in the Synchrotron Soleil data repository. Examples can be provided if needed.

\section*{Code availability}
The PYNX code [http://ftp.esrf.fr/pub/scisoft/PyNX/doc/] used for ptychographic reconstruction and the aXis2000 program [http://unicorn.mcmaster.ca/aXis2000.html] used to generate images and spectra from the reconstruction results are available for free from the indicated websites.



%

\end{document}